\documentstyle[twoside,fleqn,espcrc2]{article}

\newcommand{\AmS}{{\protect\the\textfont2
  A\kern-.1667em\lower.5ex\hbox{M}\kern-.125emS}}

\title{Absence of physical walls in hot gauge theories}

\author{Joe Kiskis
\address{
Department of Physics,
University of California, Davis, CA 95616, USA}%
\thanks{email: jekiskis@ucdavis.edu.
This research was supported by the United States Department of Energy.}}
       
\begin{document}

\begin{abstract}
This paper shows that there are no {\em physical} walls
in the deconfined, high-temperature phase of $Z(2)$ lattice
gauge theory.
In a Hamiltonian formulation, the interface in the Wilson lines is not
physical. The line interface and its energy are interpreted in terms of
physical variables.
They are associated with a difference between two partition
functions. One includes only the configurations with even flux across the
interface. The other is restricted to odd flux.
\end{abstract}

\maketitle
%
%
%%%%%%%%%%%%%%%%%%%%%%%%%%%%%%%%%%%%%%%%%%

\section{INTRODUCTION}

This paper discusses some aspects of the global $Z(N)$ symmetry \cite{r1} of
finite-temperature gauge theory.
It contributes to recent discussions of the physics of
$Z(N)$ phases and interfaces \cite{r2,me,r2.1,r2.2,r2.3,r2.4,r6,r8,walls}.
The example of $Z(2)$ gauge theory is treated here.

The Wilson line carries a nontrivial representation of the global
$Z(N)$ symmetry. In the confining phase, $\langle L \rangle=0$, and the
ensemble
is $Z(N)$ symmetric. In the high $T$ phase, $\langle L \rangle$ takes one of
$N$ distinct values proportional to the $Nth$ roots of unity $z$ in $Z(N)$, 
and the $Z(N)$ symmetry is broken.

In the Hamiltonian description, 
the physical variables are the group elements on the links of the spatial
lattice.
In a Lagrangian formulation, there are also group
elements on links in the inverse-temperature direction.
These are unphysical, auxiliary variables introduced
to enforce the Gauss law constraints. The Wilson line is constructed from the
unphysical variables.
It is a projection operator that  forces the gauge field to be in a
fundamental rather than a singlet state at the spatial position of the line.
The global $Z(N)$ symmetry of the Lagrangian formulation is not physical; it
acts as the identity on all physical states \cite{me,r2.1,r2.2,r2.4}. 
There is a single physical, high-temperature phase, which is the same for all
$z$.

If there were $N$, degenerate, physically-distinct,
high-temperature phases, then there could be large regions of space in
different phases and separated by interfaces. These walls could have a
physically-significant interface tension. In the line variables, there are
phase separations and interfaces. However,  as
noted above, the $N$ phases of the Wilson lines are not physically distinct.
Thus one may question the physical relevance of the line interfaces.

It is shown below that there are no
interfaces in the physical variables associated with the interfaces in the
lines. The physical effect associated with the interfaces in the line variables
is the ``conservation'' $mod$~2 of
the total flux along the direction perpendicular
to the line interface. The interface tension for the lines is related to
a difference between two partition
functions. One includes only the configurations with even flux across the
interface. The other is restricted to odd flux.

Because there are no physically distinct high-temperature phases
of the Yang-Mills field, there are also no metastable states in the presence
of matter. Matter simply lowers the free energy. It is shown that this
can be understood in terms of the physical variables as an increase in entropy.
       
\section{NO PHYSICAL WALLS IN HOT YANG-MILLS THEORY}

This section treats gauge theory without matter fields. The Hamiltonian
formulation in the $A_0=0$ gauge is used. The physical degrees of freedom are
gauge group elements on links of the spatial lattice.
The flux basis specifies that the amplitude to be at different points
in the group is given by a matrix element from an irreducible
representation. Physical states are configurations of physical variables with
the additional condition that the Gauss law constraints are satisfied. 
In the case of
$Z(2)$, the flux is either zero or one on each link. The constraints are that
each site must have an even number of links with ones.
To enforce these constraints, the
unphysical variables are introduced in the Lagrangian formulation. These are
the group elements on links in the fourth direction or the fourth component of
the gauge field. 

If the partition function is written as a sum over physical states, then the
unphysical variables have no work to do and need not even appear. Clearly,
interfaces in the unphysical variables are not physical per se. So the question
is: do they reflect the existence of interfaces in the physical
variables?

Previous work on the high-temperature, homogeneous phases of the
lines \cite{me,r2.1,r2.2} makes the existence of physical interfaces seem very
unlikely. It was shown that the physical configurations
associated with the $N$ pure phases of the Wilson lines at high temperature are
the same. Thus, in terms of physical variables, there is a single
high-temperature ensemble. There are not $N$ physical phases that could be
separated by walls.

The analysis uses the $Z(2)$ Ising flux model \cite{me}.
This model is equivalent to the $Z(2)$ spin Ising model. It is also
an approximation to $Z(2)$ gauge theory. Thus, the discussion and conclusions
also apply to $Z(2)$ gauge theory.
In this model, space is a three-dimensional cubic lattice. 
There are variables $\theta$ on links that can have the values 1 or 0 to
indicate the presence or absence of flux.
These are the physical variables.
A configuration is specified by the function $\theta(l)$. States with definite
flux $|\theta\rangle$ are labeled by that function. 
A general state has a wave
functional that gives the amplitude for the system to be found in the various
basis states of definite flux.
The energy of a link with flux is
$\sigma$. 
The sum over configurations is restricted to those in which the number of links
at a site that have flux is even. Let the collection of all such
configurations be $C'$. It is a subset of the unrestricted collection
of all configurations $C$. The partition function is
\begin{eqnarray}
 Z & = & Tr'[e^{-H/T}] \equiv \sum_{C'} 
                     \langle \theta | e^{-H/T} | \theta \rangle 
                                                   \label{e101}  \\
    & = & \sum_{C'} e^{ - (1/T) {\displaystyle \Sigma_{l} 
                                  \sigma \theta(l)}} .
                                                              \label{e200}
\end{eqnarray}

In a gauge theory, the Hamiltonian comes from the transfer matrix. There is a
factor of this form and a factor from the spatial plaquette term in the
Lagrangian.  It is sufficient to consider the high-temperature and
strong-coupling limit. In that case, the plaquette term can be neglected in a
first approximation.
The equivalence of this flux model to the Ising model
results from using site variables to enforce the restriction
on configurations, and then doing the $\theta$ sums.
The site variables are the Ising spins. 
First, consider the sum of $\theta(l)$ 
over the $2d=6$ $l$'s contained in the set $I(i)$ of links with endpoint $i$:
\begin{equation}
  \Sigma(i) \equiv \sum_{l \in I(i)} \theta(l)  .
\end{equation}
To force this to be even at each site, introduce the site variables $s(i)$
that take the values $\pm 1$. The factor
\begin{equation}
 \frac{1}{2} \sum_{s(i)=\pm 1} s(i)^{\Sigma(i)}            \label{e102}
\end{equation}
has the desired effect. A factor like this is introduced into the
partition function sum for each site $i$. The spins $s(i)$ are the same as the
Wilson lines of the $Z(2)$ gauge theory. This model is equivalent to the Ising
model for the spins $s(i)$ with the Ising $\beta$ related to the gauge $T$ by
$e^{-\sigma/T}=\tanh \beta$. 
Note that I never refer to the Ising model $1/\beta$ as temperature.
Temperature always refers to the $T$ appearing in (\ref{e101}).
Large $T$ gives large $\beta$, and ordered Ising spins.
It is convenient to introduce $\mu \equiv \sigma / T$.
In the high $T$ region,
\begin{equation}
  e^{-2\beta} \sim \mu/2  .           \label{e80}
\end{equation}

To discuss interfaces, it helps to introduce boundary conditions that force at
least one interface in the lines to appear. Use
solid-cylindrical geometry: space is finite in the transverse $x$ and $y$
directions and very long in the
$z$ direction. 
There are periodic boundary conditions in the $x$ and $y$ directions. The
transverse area in dimensionless lattice units is $A$.
Let $L$ be large and positive.
Apply the boundary conditions $s=-1$ for $z = -L/2$ and $s=1$ for $z = L/2$.
The limit of interest is $L \rightarrow \infty$ followed by 
$A \rightarrow \infty$.

First review the picture in terms of the unphysical spin variables.
The high-temperature region is 
the large-$\beta$, ordered phase for the Ising spins.
The interface is approximately flat and
has a Boltzmann weight $e^{-2\beta A}$.
Thus, the energy per unit area $\alpha$ is related to $\beta$
by $\alpha/T \approx 2\beta$ as $\beta \rightarrow \infty$.
The partition function with the indicated boundary conditions is $Z'$ and is a
sum with an odd number of interfaces.
The partition function with the same boundary conditions at each end 
and an even number of interfaces is $Z$.
The ratio is
\begin{equation}
  Z'/Z  =  e^{-(F'-F)/T}  \cong  1 - (F'-F)/T  . 
\end{equation}
In the large $L$ limit, the result of the sums is the excess free energy 
\begin{equation}
 F'-F = 2 T e^{-2 L \epsilon } \; \; {\rm with } \; \;  
             \epsilon = e^{-2\beta A}.            \label{e21}
\end{equation}
This relates the activity for the wall
$e^{-2\beta A}$ to $Z'/Z$ and $F'-F$. The approximations are valid for
large $L$ and $Le^{-2\beta A} \gg 1$ .

Now consider the same situation in terms of flux variables.
In the high-temperature, deconfined phase with $\mu= \sigma/T$ small,
there is dense, percolating flux.
The flux is almost random except for the
constraint that there be an even number of links with flux at each site.

To enforce the constraints, there is a factor given in (\ref{e102}) 
at each site. For a site on the $z=-L/2$ boundary, this becomes
$(-1)^{\theta}$.
This gives a factor of
$-1$ for each link with flux coming into the system from the boundary. 

From the local Gauss law constraints, it follows that
the total flux $\sum_{xy} \theta_z(x,y,z)$ $mod$ 2 on a transverse slice of
longitudinal links is independent of $z$.
This will be referred to as flux ``conservation''. Thus each configuration can
be described as
even or odd. The effect of the boundary conditions is to weight the odd
configurations with and extra factor of $-1$. The partition function sum
\begin{equation}
 Z = \sum_{even} e^{-H/T} + \sum_{odd} e^{-H/T}
\end{equation}
is replaced by
\begin{equation}
 Z' = \sum_{even} e^{-H/T} - \sum_{odd} e^{-H/T} \label{e201}  .
\end{equation}
This gives
\begin{eqnarray}
  Z'/Z & = & (Z_e - Z_o)/(Z_e + Z_o) \\
       & = & 1-2e^{-(F_o-F_e)/T} .  \label{e20}
\end{eqnarray}
The last step is correct for large $(F_o-F_e)$.
This free energy difference is 
proportional to the
length $L$, so that (\ref{e20}) is consistent with (\ref{e21}). 
Thus, the free energy difference per unit length of odd flux verses even flux
is the quantity of interest.

The next part of the discussion shows the direct connection between the spin
interfaces and the conservation of flux.
Consider the structure of the partition function.
The role of the spins is to enforce the local constraints on flux. 
In the calculation leading to (\ref{e21}),
the spins are constant on transverse planes. Thus, they are enforcing the
weaker constraint of flux conservation.
The calculation begins from (\ref{e200}), inserts the factors (\ref{e102}),
does the
flux sum, and then does the spin sum. It is manifest in the calculation that
the interfaces enforce the flux conservation.
The result is \cite{walls}
\begin{equation}
  (F_o-F_e)/T = 2L(\frac{\mu}{2})^A  .
\end{equation}
Since $\mu$ and $\beta$ are related by (\ref{e80})
this is equivalent to (\ref{e21}).

This shows that the interfaces in the unphysical
spin variables are associated with flux conservation i.e.\ the
total flux $mod$ 2 on a transverse slice of links is independent of
longitudinal position. 

With that established, one can redo the calculation of $Z'$ by a much easier
method using flux variables only \cite{walls}. The partition function $Z'$ is
the difference of two contributions. It is the sum of all the even
configurations minus the sum of all the odd configurations as in (\ref{e201}).
There is no reference to interfaces. The result is the same as that
from the spin calculation, but the associated physical picture is completely
different.  In flux, it is the Gauss law constraints rather than interfaces
that are important.

\section{EFFECTS OF MATTER}

Matter fields in the
fundamental representation explicitly break the global $Z(N)$ symmetry.  The
$N$ line states are no longer degenerate. Above the critical temperature, there
can be metastable states of the line variables separated by 
walls \cite{r8}.

The important effect of the matter is that it breaks the
global symmetry.
This can be studied in a simpler situation where 
the matter is represented by an external field $h$ linearly coupled to the
Wilson lines: $h^* L + h L^* $.

In a general situation with $Z(2)$ global symmetry, the constraint effective
potential with $h \neq 0$ is a double well with one side lower than the other.
As $h \rightarrow 0$, this becomes the symmetric double well with degeneracy.
Thus the existence of the metastable state is linked to the existence of
degeneracy.

However, without matter, there is a single, physical, high-temperature phase.
Thus, there is no degeneracy that could be connected to multiple states,
some metastable, when symmetry breaking is present. I conclude that
in the physical flux picture, there are no
metastable states.

Nevertheless, matter does affect the free energy. It is interesting to
describe this effect entirely in the flux picture. 
When the partition function is expressed in
terms of flux variables, the effect of $h$ is to allow odd sites with a
density controlled by $h$. This increases the number of
allowed configurations (i.e.\ increases the entropy), 
increases $Z$, and decreases $F$.

\section{CONCLUSION}

While calculations in terms of the
physical flux variables and in terms of the unphysical lines lead to the same
result, the associated interpretations are completely different.
Interfaces in the line variables may be a
convenient device for making approximate calculations of physical quantities in
terms of unphysical variables. However, one should be cautious with heuristic
arguments that rely upon the physical reality of the interfaces.
In terms of physical variables, there are no interfaces or metastable states.

\end{document}